\documentclass[aps,prl,reprint,groupedaddress,showpacs]{revtex4-1}
\usepackage{graphicx}% Include figure files
\usepackage{dcolumn}% Align table columns on decimal point
\usepackage{bm}% bold math
\usepackage[latin1]{inputenc}

\begin{document}

\title{Nonlinear optical response of a two-dimensional atomic crystal}

\author{Michele Merano}
\email[]{michele.merano@unipd.it}

\affiliation{Dipartimento di Fisica e Astronomia G. Galilei, Universit$\grave{a}$ degli studi di Padova, via Marzolo 8, 35151 Padova, Italy}

\date{\today}

\begin{abstract}
The theory of Bloembergen and Pershan for the light waves at the boundary of nonlinear media is extended to a nonlinear two-dimensional atomic crystal, i.e. a single planar atomic lattice, placed in between linear bulk media. The crystal is treated as a zero-thickness interface, a real two-dimensional system. Harmonic waves emanate from it. Generalization of the laws of reflection and refraction give the direction and the intensity of the harmonic waves. As a particular case that contains all the essential physical features, second order harmonic generation is considered. The theory, due to its simplicity that stems from the special character of a single planar atomic lattice, is able to elucidate and to explain the rich experimental details of harmonic generation from a two-dimensional atomic crystal. 
\end{abstract}

\maketitle
\section{}

A two-dimensional (2D) atomic crystal consists of a single planar atomic lattice. These materials, were shown to exist and to be stable by marvelous experiments \cite{Novoselov2004, Novoselov2005}. Few-layer materials, consisting of a few planar atomic lattices, can be created as well and they are also interesting. Anyway single layer materials are particularly special. For instance the overlapping between the valence and the conduction band in graphene is exactly zero, while it is finite in graphite. Single-layer transition metal dichalcogenides are direct band semiconductors while the bulk materials (or the few-layer ones) have an indirect band gap \cite{Heinz2010}.

Also the linear optical properties of a 2D atomic crystal are remarkable. The fine structure constant defines the optical transparency of graphene \cite{Nair2008}. Thanks to an enhanced optical contrast, 2D atomic crystals can be well visualized if deposited on top of suitable substrates \cite{Blake2007, Kis11, Sandhu13}. In two recent papers it was shown that for light a single-layer material has no thickness \cite{Merano15, Merano215}. In practice for optics it appears as a real 2D system. This result is by no means obvious. When 2D atomic crystals are deposited on a substrate, atomic force microscopy tips can be used to measure their thickness \cite{Novoselov2004, Novoselov2005}. But a model treating them as three dimensional (3D) slabs with a certain thickness fails to explain the overall experiments on linear light matter interaction (absorption for instance) \cite{Merano15}. Instead a model considering the 2D atomic crystal as part of the interface plus the right boundary conditions, turns out to be the successful approach to explain its linear optical properties \cite{Merano15}.

Nonlinear optical properties of single layer and few layer atomic crystals have been investigated recently and they exhibit an extremely rich variety of physical phenomena. Single-layer $\rm MoS_{2}$ \cite{Zhao13, Paula13, Heinz13, Kim14} and BN \cite{Heinz13} are non centrosymmetric materials, while their bilayers and bulk counterparts are expected to exhibit inversion symmetry \cite{Paula13, Heinz13, Kim14}. More specifically slices with an even number of layers belong to the centrosymmetric $D_{3d}$ space group, while slices with an odd number of layers belong to the non-centrosymmetric $D_{3h}$ space group. Strong second harmonic generation (SHG) from materials with an odd number of layers was observed while the centrosymmetric materials do not yield appreciable SHG \cite{Zhao13, Paula13, Heinz13, Kim14}. The efficiency of SHG shows a dramatic even-odd oscillation with the number of layers even for $\rm WS_{2}$ and $\rm WSe_{2}$ \cite{Cui13}. It was also shown that the SHG depends on the 2D atomic crystal orientation with respect to the polarization of the fundamental incident beam, revealing in this way the rotational symmetry of the lattice \cite{Zhao13, Paula13, Heinz13, Kim14}. Because the ideal free standing monolayer graphene is centrosymmetric, its SHG vanishes within the dipole approximation \cite{Osgood13}. In contrast symmetry-allowed third order nonlinear optical effects in graphene are remarkably strong \cite{Zhao213, Osgood13, Mikhailov10, Zhang12}. 

The nonlinear optical response of single layer and few layer atomic crystals is described by modelling them as 3D slabs of homogeneous media with a certain thickness and a certain refractive index \cite{Osgood13, Zhao13, Kim14, Zhao213}. This model proved to be successful in different cases but since it fails to explain the overall linear optical properties of a 2D atomic crystal, i.e. a single planar atomic lattice \cite{Merano15}, its use, at least for this class of materials, is not completely satisfactory.

Here I show that, in analogy to what happens for linear optics, it is possible to describe the nonlinear optical response of a 2D atomic crystal by modelling it as a zero-thickness interface plus the right boundary conditions. Like for bulk materials an example that contains all the essential physical features is provided by SHG \cite{Bloembergen62}. An experiment is analyzed to show the strength of this approach.

Consider a monochromatic plane wave at frequency $\omega$ incident on a plane 2D atomic crystal which lacks inversion symmetry. The crystal is deposited at the interface of two linear bulk media (Fig. 1). The nonlinear material properties are described by expanding the surface density of polarization $\vec{\textbf{\emph{P}}}$ in a power series in the field. For the pure electric dipole case one has \cite{Baldwin}:
\begin{eqnarray}
\vec{\textbf{\emph{P}}}=\epsilon_{0} \vec{\chi} \cdot \vec{\textbf{\emph{E}}}+\epsilon_{0} \vec{\chi_{2}}: \vec{\textbf{\emph{E}}}\vec{\textbf{\emph{E}}}+...
\end{eqnarray}
where $\vec{\textbf{\emph{E}}}$ is the total electric field in the 2D atomic crystal, $\vec{\chi}$ is a second rank tensor and $\vec{\chi_{2}}$ is a third rank tensor. The first term defines the usual surface linear susceptibility, the second term, the lowest surface nonlinear susceptibility, and so on. This procedure is useful because the optical nonlinearities are small \cite{Bloembergen}. This has already been confirmed by experiments even in the case of 2D atomic crystals \cite{Zhao13, Heinz13, Paula13, Kim14, Cui13, Osgood13, Zhao213, Mikhailov10, Zhang12}. The nonlinear susceptibility $\vec{\chi_{2}}$ of the medium will give rise to a polarization at the second harmonic frequency $\omega_{2}=2\omega$ that will act as a nonlinear source of electromagnetic plane waves.

\begin{figure}
\includegraphics{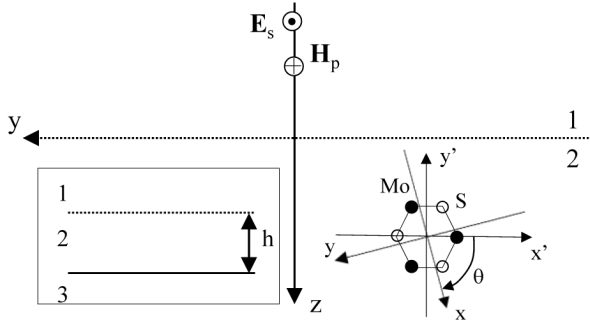}
\caption{A plane wave (not shown) is normally incident on a $\rm MoS_{2}$ monolayer deposited on a substrate (medium 2). The second harmonic generation is investigated. The electric (magnetic) field for $s$ ($p$) polarization at $\omega$ and $2\omega$ is shown. Left inset: 3 layer substrate; $\rm MoS_{2}$ is deposited at the interface of bulk media 1 and 2. Right inset: orientation of crystallographic axes (x'. y', z') of a $\rm MoS_{2}$ monolayer with respect to laboratory axes (x, y, z). The orientation of BN is the same with B replacing Mo and N at the place of the projection of the $\rm S_{2}$ atoms on the x'y' plane (see ref. \cite{Heinz13}).}
\end{figure}

The boundary conditions are $\hat{\kappa} \wedge (\vec{\textbf{E}}_{2}-\vec{\textbf{E}}_{1})=0$, $\hat{\kappa} \wedge (\vec{\textbf{H}}_{2}-\vec{\textbf{H}}_{1})=\textbf{J}_{p}$ where $\hat{\kappa}$ is the unit vector along the z axis and $\vec{\textbf{J}}_{p}=\partial{\vec{\emph{\textbf{P}}}}/\partial{t}$ \cite{Merano15}. The tangential components of $\vec{\textbf{\emph{E}}}$ and $\vec{\textbf{\emph{H}}}$ must satisfy these conditions on the boundary everywhere and at all times. This requires that the individual frequency components, at $\omega$ and $2\omega$, are separately continuous across the boundary \cite{Bloembergen62}. To satisfy these conditions one requires for the fundamental frequency:
\begin{eqnarray}
k_{1yi}=k_{1yr}=k_{1yt}
\end{eqnarray}
where $k_{1yi}$, $k_{1yr}$, $k_{1yt}$ are the $y$ components of the wavevectors for the incident, reflected and transmitted wave. For the $y$ components of the wavevectors at $2\omega$:
\begin{eqnarray}
k_{2yr}=k_{2yt}.
\end{eqnarray}
Obviously the second harmonic plane waves only propagate in the direction of the reflected and transmitted fundamental plane waves.  These relations reflect the general requirement of conservation of the tangential component of momentum \cite{Bloembergen62}.

To avoid unnecessary complications I develop formulas for the experimental conditions of the papers that will be later considered (Fig. 1). First of all normal incidence, with $s$ polarization, the generalization to an angle of incidence different from zero is not more difficult than in the linear case. I already specify the symmetry of the crystal used in experiments being that of single-layer BN and $\rm MoS_{2}$ (Fig1, right inset). From refs \cite{Zhao13, Heinz13} the $D_{3h}$ symmetry implies that the non zero elements of $\vec{\chi_{2}}$ are $\chi_{2y'y'y'}=-\chi_{2y'x'x'}=-\chi_{2x'x'y'}=-\chi_{2x'y'x'} =\chi_{2}$, where $x'$ $y'$ $z'$ are crystallographic coordinates and $z'$ is opposite to the incident plane wave direction $z$. I also fix the fundamental incident wavelength $\lambda$ = 810 nm equal to that used in \cite{Zhao13, Heinz13}. This wavelength corresponds to an incident energy photon that is minor that the energy band gap of both the BN and the $\rm MoS_{2}$ \cite{Heinz2010, Blake2011}. In this case, it is safe to consider the surface conductivity ($\sigma$) of these materials equal to zero. These crystals are non magnetic so I assume a null surface magnetic susceptibility. As in ref \cite{Merano15} I assume for $\vec{\chi}$ in-plane isotropy and null out-of-plane values.

The relation in between $\vec{\emph{\textbf{E}}}$ and $\vec{\emph{\textbf{H}}}$ in the incident, reflected and transmitted waves is:
\begin{eqnarray}
\frac{\eta}{n_{1}}\vec{\textbf{\emph{H}}}_{i(r)}=\hat{s}_{i(r)}\wedge\vec{\textbf{\emph{E}}}_{i(r)}; \quad \frac{\eta}{n_{2}}\vec{\textbf{\emph{H}}}_{t}=\hat{s}_{t}\wedge\vec{\textbf{\emph{E}}}_{t};
\end{eqnarray}
where $n_{1}$, $n_{2}$ are the refractive indexes. Of course $n_{1}$, $n_{2}$ are different at $\omega$ and $2\omega$. The boundary conditions at $\omega$ run:
\begin{eqnarray}
E_{1xi}+E_{1xr}&=&E_{1xt}\\
E_{1xi}+E_{1xr}&=&\frac{P_{1x}}{\epsilon_{0}\chi} \nonumber  \\
H_{1yi}-H_{1yr}&=&H_{1yt}+i\omega P_{1x} \nonumber
\end{eqnarray}
showing that at this frequency the directions and the amplitudes of the reflected and the refracted waves are determined by the Fresnel laws for the linear medium \cite{Bloembergen62}. The wave at $2\omega$ has an $s$ polarized component given by:
\begin{eqnarray}
P_{2x}&=&\epsilon_{0} \chi_{2}\sin(3\theta)(E_{1xi}+E_{1xr})^2 \\
E_{2xr}&=&E_{2xt} \nonumber\\
-H_{2yr}&=&H_{2yt}+2i\omega P_{2x} \nonumber
\end{eqnarray}
and a $p$ polarized component given by:
\begin{eqnarray}
P_{2y}&=&\epsilon_{0} \chi_{2}\cos(3\theta)(E_{1xi}+E_{1xr})^2 \\
-E_{2yr}&=&E_{2yt} \nonumber\\
H_{2xr}&=&H_{2xt}+2i\omega P_{2y} \nonumber
\end{eqnarray}
Here $\theta$ is the angle between the $x$ laboratory coordinate and the $x'$ crystallographic coordinate (Fig. 1, right inset). From (6) and (7):
\begin{eqnarray}
&E_{2xr}&=\frac{-2i\omega \eta}{n_{1}+n_{2}}P_{2x}=-ik_{2}\chi_{2}\sin(3\theta)\frac{(1+r_{s})^2}{n_{1}+n_{2}}E^2_{1xi}  \\
&E_{2yr}&=\frac{2i\omega \eta}{n_{1}+n_{2}}P_{2y}=ik_{2}\chi_{2}\cos(3\theta)\frac{(1+r_{s})^2}{n_{1}+n_{2}}E^2_{1xi} \nonumber
\end{eqnarray}
where $r_{s}$ is the reflection coefficient for the wave at $\omega$ given in ref. \cite{Merano15} and $n_{1}$, $n_{2}$, $k_{2}$ are the refractive indices in media 1 and 2, and the wavevector in vacuum for the wave at $2\omega$. For $n_{1}=n_{2}=1$ we retrieve the case for a free standing layer.

The SHG of a 2D atomic crystal on a stratified medium is of primary importance also. I consider the case represented in Fig. 1 (left inset), corresponding to published experimental work that will be considered in this article. The boundary conditions for the interface in between bulk media 2 and 3 are $\hat{\kappa} \wedge (\vec{\textbf{E}}_{3}-\vec{\textbf{E}}_{2})=0$, $\hat{\kappa} \wedge (\vec{\textbf{H}}_{3}-\vec{\textbf{H}}_{2})=0$. Equation (4) is easily extended to medium 3. Again the $s$ polarized fundamental wave obeys the same equations for the linear case \cite{Merano15}:
\begin{eqnarray}
E_{1xi}+E_{1xr}&=&E_{1x+}+E_{1x-} \\
E_{1xi}+E_{1xr}&=&\frac{P_{1x}}{\epsilon_{0}\chi} \nonumber  \\
H_{1yi}-H_{1yr}&=&H_{1y+}-H_{1y-}+i\omega P_{1x} \nonumber \\
E_{1xt}&=&E_{1x+}e^{-i\beta_{1}}+E_{1x-}e^{i\beta_{1}} \nonumber\\
H_{1yt}&=&H_{1y+}e^{-i\beta_{1}}-H_{1y-}e^{i\beta_{1}} \nonumber
\end{eqnarray}
where $\beta_{1}=k_{1}n_{2}h$ and $h$ is the thickness of medium 2. The $s$ polarized wave at $2\omega$ satisfies:
\begin{eqnarray}
P_{2x}&=&\epsilon_{0} \chi_{2}sin(3\theta)(E_{1xi}+E_{1xr})^2\\
E_{2xr}&=&E_{2x+}+E_{2x-}; \nonumber \\ -H_{2yr}&=&H_{2y+}-H_{2y-}+2i\omega P_{2x} \nonumber \\
E_{2xt}&=&E_{2x+}e^{-i\beta_{2}}+E_{2x-}e^{i\beta_{2}}\nonumber \\ 
H_{2yt}&=&H_{2y+}e^{-i\beta_{2}}-H_{2y-}e^{i\beta_{2}} \nonumber
\end{eqnarray}
and for the $p$ polarized wave at $2\omega$ we have:
\begin{eqnarray}
P_{2y}&=&\epsilon_{0} \chi_{2}\cos(3\theta)(E_{1xi}+E_{1xr})^2\\
-E_{2yr}&=&E_{2y+}-E_{2y-}; \nonumber \\ 
H_{2xr}&=&H_{2x+}+H_{2x-}+2i\omega P_{2y} \nonumber \\
E_{2yt}&=&E_{2y+}e^{-i\beta_{2}}-E_{2y-}e^{i\beta_{2}} \nonumber \\ 
H_{2xt}&=&H_{2x+}e^{-i\beta_{2}}+H_{2x-}e^{i\beta_{2}} \nonumber
\end{eqnarray}
where $\beta_{2}=k_{2}n_{2}h$. I obtain:
\begin{eqnarray}
E_{2xr}&=&\frac{-ik_{1} \chi_{2}}{n_{1}}\sin(3\theta)(1+r_{1C23s})^2(1+r_{123s})E^2_{1xi} \qquad   \\
E_{2yr}&=&\frac{ik_{1} \chi_{2}}{n_{1}}\cos(3\theta)(1+r_{1C23s})^2(1+r_{123s})E^2_{1xi} \nonumber
%&E&_{2xt}=-ik_{1} \chi_{2}\sin(3\theta)(1+r_{1C23s})^2 t_{123s}E^2_{1xi}      \nonumber
\end{eqnarray}
%\begin{eqnarray}
%&E&_{2yr}=\frac{ik_{1} \chi_{2}}{n_{1}}\cos(3\theta)(1+r_{1C23s})^2(r_{123p}-1)E^2_{1xi} \quad \\
%&E&_{2yt}=-\frac{ik_{1} \chi_{2}}{n_{1}}\cos(3\theta)(1+r_{1C23s})^2 (t_{123p})E^2_{1xi}\nonumber
%\end{eqnarray}
where $r_{1C23s}$ is the reflection coefficient for the $s$ wave at $\omega$ given in \cite{Merano15} (C stays for 2D atomic crystal), and $r_{123s}$ is the $s$ wave reflection coefficient for the bare substrate at $2\omega$.
\begin{figure}
\includegraphics[width=8 cm]{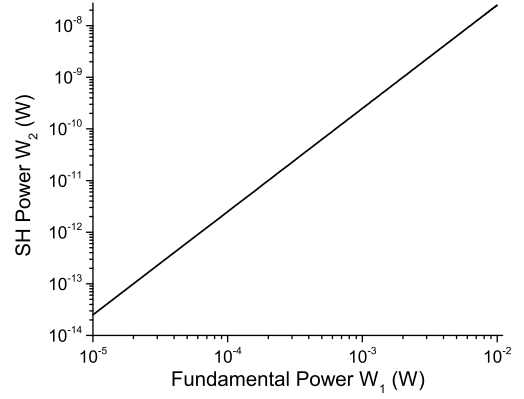}
\caption{Simulation of the experimental results \cite{Zhao13} for the second harmonic generation from a $\rm MoS_{2}$ mono-layer deposited onto a $\rm Si/SiO_{2}$ (90 nm) substrate. The reported power dependence is based on formula (13) for a $\chi_{2} =2\cdot 10^{-18}$ $\rm m^{2}/V$.}
\end{figure}

Among the papers on SHG from 2D atomic crystals I will focus mainly on ref. \cite{Zhao13} that reports SHG intensities directly comparable with theory and measurements for two different substrates. Other experiments published in \cite{Heinz13, Paula13, Kim14} will be a valuable support as well. From (12) I retrieve the fact that the $s$ and the $p$ reflected components of the harmonics are proportional to $\sin (3\theta)$ and $\cos (3\theta)$. This was proven to be valid for $\rm MoS_{2}$ \cite{Zhao13, Paula13, Heinz13, Kim14} and for BN \cite{Heinz13} as well, showing the remarkable result that SHG provides a purely optical method of determining the orientation of the crystallographic axes. At normal incidence, the total power is independent of $\theta$ \cite{Zhao13, Paula13, Heinz13, Kim14}. In ref. \cite{Zhao13} authors study single layers $\rm MoS_{2}$ deposited onto $\rm Si/SiO_{2}$ (90 nm) and onto $\rm Si/SiO_{2}$ (280 nm) substrates at a fixed fundamental wavelength (810 nm).  The incident pulse energy is gaussian in both time and space with full widths at half maxima of $\tau$ = 200 fs and $w$ = 2 $\mu$m. The laser repetition rate is $f$ = 81 MHz. Assuming the same parameters for the wave at $2\omega$, its total reflected power $W_{2}$ is related to the input fundamental power $W_{1}$ via:
\begin{eqnarray}
&W_{2}&=(\mid E_{2xr}\mid^2 + \mid E_{2yr}\mid^2)\left(\frac{2\sqrt{2ln2}}{\sqrt{2 \pi }}\right)^3 \frac{W^2_{1}}{\epsilon_{0}f w^2 c\tau}=  \qquad  \\
&=&\mid\frac{ik_{1} \chi_{2}}{n_{1}}(1+r_{1C23s})^2(1+r_{123s})\mid^2 \left(\frac{2\sqrt{2ln2}}{\sqrt{2 \pi }}\right)^3 \frac{W^2_{1}}{\epsilon_{0}f w^2 c\tau} \quad \nonumber
\end{eqnarray}

Figure 2 shows $W_{2}$ as a function of $W_{1}$ for a $\rm MoS_{2}$ monolayer on a  $\rm Si/SiO_{2}$ (90 nm) substrate \cite{Zhao13}. The curve furnishes a $\chi_{2} = 2\cdot 10^{-18}$ $\rm m^2/V$. This value is similar to the one reported ($\chi_{2}=10^{-19}$ $\rm m^2/V$) in \cite{Heinz13} at the same fundamental wavelength. While the experiment reported in \cite{Zhao13} is certainly very well done, the data analysis providing the $\chi_{2}$ values must contain an error since the reported $\chi_{2}$ are order of magnitude bigger than expected if compared with my analysis and with ref. \cite{Heinz13}. Other authors measured SHG from a $\rm MoS_{2}$ monolayer \cite{Paula13, Kim14}, they report values of $\chi_{2}$ similar to \cite{Heinz13}. Reference \cite{Kim14} explicitly claims that their experimental intensities are similar to those in \cite{Zhao13}. Theoretically $W_{2}$ depends also on the linear susceptibility $\chi$ (because of $r_{1C23s}$). From the experimental data in ref. \cite{Heinz2014}, where the reflectance as a function of the incident wavelength of a $\rm MoS_{2}$ monolayer deposited on a fused silica substrate is reported, and formulas in \cite{Merano15}, I obtain $\mid \chi \mid<2 \cdot 10^{-8}$ m at 810 nm. From simulations, such small values of $\chi$ are irrelevant on $W_{2}$ and not distinguishable from the case at $\chi =0$.

Figure 3 shows $W_{2}$ as a function of $W_{1}$ for a $\rm MoS_{2}$ monolayer on a  $\rm Si/SiO_{2}$ (280 nm) substrate \cite{Zhao13}. The curve furnishes a $\chi_{2} = 10^{-18}$ $\rm m^2/V$. The difference with the previous case is only a factor 2. This is in contrast with ref. \cite{Zhao13} where authors estimate a $\chi_{2}$ 20 times smaller than in the previous case. Authors attribute this discrepancy to the fact that in the first case the $\rm MoS_{2}$ monolayer is mechanically exfoliated and in the second case is grown by CVD. This may be possible. But the linear properties of exfoliated and CVD $\rm MoS_{2}$ flakes are not different at this wavelength (see data in ref. \cite{Heinz2014}). Reference \cite{Kim14} also comments on this difference saying that for CVD-grown samples they find the same $\chi_{2}$ value as for the exfoliated ones (at different wavelenghts from those consideres in \cite{Zhao13}.)
\begin{figure}
\includegraphics[width=8 cm]{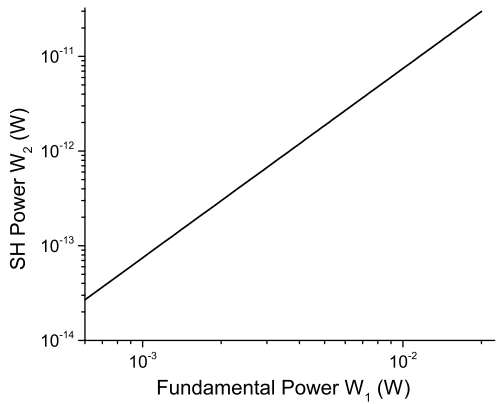}
\caption{Simulation of the experimental results \cite{Zhao13} for the second harmonic generation from a $\rm MoS_{2}$ mono-layer deposited onto a $\rm Si/SiO_{2}$ (280 nm) substrate. The reported power dependence is based on formula (13) for a $\chi_{2} = 10^{-18}$ $\rm m^{2}/V$.}
\end{figure}

The difference in between the SHG intensities from a $\rm MoS_{2}$ monolayer on $\rm Si/SiO_{2}$ (90 nm) and on $\rm Si/SiO_{2}$ (280 nm) substrates has a simple explanation. The second order polarization in the experiment here analyzed is proportional to $(E_{1xi}+E_{1xr})^2$. If I consider the $\rm MoS_{2}$ monolayer as part of the interface, as a real 2D system, the reflection coefficient fixing $E_{1xr}$ depends strongly on the substrate on which it is deposited. From formula (13) $W_{2}\propto \mid (1+r_{1C23s})^2 (1+r_{123s})\mid ^2$. For the $\rm Si/SiO_{2}$ (90 nm) substrate $\mid (1+r_{1C23s})^2 (1+r_{123s})\mid ^2= 2.1$ and for the $\rm Si/SiO_{2}$ (280 nm) substrate $\mid (1+r_{1C23s})^2 (1+r_{123s})\mid ^2=2.5\cdot 10^{-3}$. From these results it is clear that the different SHG intensities from the $\rm MoS_{2}$ flakes in ref. \cite{Zhao13} are due to the high sensitivity of the nonlinear optical response of 2D atomic crystals to the substrates on which they are deposited. Judging from the published literature on the subject \cite{Zhao13, Heinz13, Paula13, Kim14}, this aspect appears to be elusive (most probably due to the extremely rich variety of physical results that characterize harmonic generations from these materials) or has been attributed to bad quality samples.

In conclusion I have extended the theory of Bloembergen and Pershan \cite{Bloembergen62} for the light waves at the boundary of nonlinear media to the case of a 2D atomic crystal defined as a single planar atomic lattice. Reference \cite{Bloembergen62} considers harmonic generation at the interface of two bulk media and also the nonlinear plane-parallel plate. Very often 2D crystals are modelled like this, i e. as very thin (compared to the incident wavelength) 3D slabs \cite{Osgood13, Zhao13, Kim14, Zhao213}. Here I adopt an approach that explains the SHG from these materials by modelling them as zero-thickness interfaces. The boundary conditions for such an interface need to be modified with respect to those adopted in \cite{Bloembergen62}. Any hypothesis on an effective thickness of the 2D atomic crystal appears not necessary. Of course the theory here developed is not anymore valid when a double-layer or a few-layer crystal is considered. My approach is simpler than the one based on a 3D slab because it does not consider multiple reflections inside a single plane of atoms as required by modelling it as plane-parallel plate. This simplicity is due to the particularly special character of single atomic layer materials. Most of all the formulas deduced here shine light on the SHG process, by quantitatively estimate all the aspects of the SHG intensity generation as the celebrated result that SHG depends on the orientation of the crystallographic axes with respect to the input polarization \cite{Zhao13, Heinz13, Paula13, Kim14}. More specifically formulas also evidence a strong dependance of SHG on the the substrate on which these crystals are deposited. The present analysis focuses on SHG but it is not difficult to extend these formulas to cases that involve other example of waves mixing or higher order nonlinear susceptibilities. Finally it is remarkable that the overall optical properties, linear \cite{Merano15, Merano215} and nonlinear, of 2D atomic crystals are described by a model considering them as zero-thickness interfaces.

\end{document}